# Spin-precession in an adiabatically rotating electric field.

Yuri A. Serebrennikov

Qubit Technology Center

2152 Merokee Dr., Merrick, NY 11566

ys455@columbia.edu

Due to spin-orbit coupling, the adiabatic perturbation of an electron's orbital motion induced by a revolving external electric field lead to the electron spin-precession. The obtained results describe both transverse and longitudinal dynamics of an effective spin, for conical and figure-8 trajectories of a driving electric field, and elucidate the link between the quantum-mechanical, geometrical, and classical pictures of spin rotation. Within the limit of adiabatic approximation the derived formulas are valid regardless of whether eigenvectors of the total Hamiltonian of the problem are explicitly available or not, and are convenient for approximate calculations. The main expression for the time evolution of the Bloch vector has pure geometric character and is independent of the physical context.

## I. Introduction

Potential realization of spin-based quantum computer architecture, where spin orientation would represent a bit (qubit) of information, has revived interest in spin dynamics and relaxation. Due to theoretical simplicity and experimental accessibility, most work focused on the coherent spin manipulation by a time-varying *magnetic* field. The fundamental interest to this rather ordinary topic stems from the astonishing discovery that the evolution of quantum systems governed by a time-dependent Hamiltonian contains terms of pure geometrical nature[1,2]. The canonical Berry's example is that of a spin-1/2 particle in an external magnetic field of a constant magnitude adiabatically traversing a conical circuit[1]. Detailed analytical treatment of this physical model helps to realize that Abelian holonomy associated with the circuit is responsible for a great variety of real-life interference phenomena ranging from general physics and chemistry to quantum information processing.

Surprisingly, an entirely electrical control of a spin motion in a time-dependent *electric* field has not received much attention so far. This is even more surprising, if one considers the well-known fact that fermionic systems with time-reversal invariance (in zero magnetic field) possess at least two-fold Kramers degeneracy at all configurations of an external electric field potential and, hence, may be considered as natural candidates for the realization of non-Abelian holonomies[3]. The latter can be achieved by adiabatically driving a degenerate system around a closed path in the parameter space. This leads to a nontrivial unitary transformation among the degenerate eigenfunctions of the instantaneous Hamiltonian. The resulting phase shifts and transitions between quantum states allow a complete geometrical description, i.e. without indicating the schedule of a motion in the parameter space. The spin-based implementation of non-Abelian holonomic computation has been proposed for exitonic states of semiconductor quantum dots (QDs)[4].

Clearly, electric field cannot directly affect the spin of a target particle. However, spin is generally coupled with orbital momentum and, hence, through spin-orbit coupling (SOC), Stark field may direct the motion of a spin. Very recently, the gate voltage control over electron spin-precession was demonstrated in 2D semiconductor nanostructures[5][6]. These results stimulated the theoretical study of electron spin motion in a time-varying Stark field. Rashba and Efros[7] developed the theory of dynamic electron-spin manipulation by an external electric field that has a time-dependent magnitude and *fixed direction* in space. It has been shown that in semiconductor quantum wells "efficient electrical spin manipulation can be achieved through the orbital mechanism of spin coupling to the electric field" [7].

The model of spin-electric coupling considered in Ref.[7] corresponds to the one proposed by Kronig and Van Vleck and represents the physical basis of the "direct" spin-lattice relaxation mechanism associated with the stochastic modulation of SOC by lattice vibrations[8]. It is well known, however, that in zero magnetic fields this mechanism of spin-flip transitions is "singularly ineffective"[9]. This is the consequence of the celebrated Kramers' theorem on the degeneracy of fermionic Hamiltonians. An electric field that has a fixed direction in space cannot break the time-reversal symmetry and split or induce transitions between the components of a Kramers' doublet.

Rotation of a Stark field can, however, violate the T-invariance of a system and due to SOC lead to the specific form of spin-electric coupling: spin-rotation interaction[10][11]. The situation becomes especially interesting for odd number of electrons in zero magnetic fields. As will be shown below, in this case one may acquire an *entirely* electric control over the spin motion. Note that random modulation of the *orientation* of principal axes of an electric field tensor will induce the spin relaxation process[12][13] complementary to that discussed recently in Refs.[14][15][16][17].

To clarify the underlying physics and to simplify the calculations, we consider a single electron confined in a spherical QD or bound by a central potential and calculate the response of an effective spin (due to SOC, spin may not be a good quantum number) to the adiabatic perturbation of the electron's orbital motion induced by the *revolving* external electric field. This can describe, e.g., the evolution of excited *p*-like electron states in a spherical CdSe nanocrystals[18] under the influence of slowly rotating Stark field. Contrary to the *s*-like electron states in the lowest conduction-band, these discrete quantum size levels of the confined electrons possess nonzero orbital momentum, $\vec{L}$, and are degenerate with respect to the projection of the total angular momentum, $\vec{J} = \vec{L} + \vec{S}$, similar to the bound electron states in atoms. As has been shown by Rodina *et al.*[18], the eight-band effective-mass model leads to the surface-induced spin-orbit Hamiltonian, $H_{SO}(1P) = \lambda(1P)\vec{L}\cdot\vec{S}$. The corresponding electron states are described by the conduction-band spinor envelope functions that are an eigen spinors of $\vec{J}$. Obviously, for *s*-like electron states in spherical QDs the matrix elements of the SOC operator are zero in the first order that does not engage the excitation of virtual states with orbital momentum quantum number $l \geq 1$. It is well known, however, that even a small admixture of terms with $l \neq 0$ to the states with $l = 0$ will couple the spin and spatial degrees of freedom and lead, e.g., to anisotropy of electron g-tensor in the lowest conduction band.

In this paper, we shall find the complete analytical solution of the time-dependent Schrödinger-type equation that describes the conical and figure-8 trajectories of a driving

Stark field. Our results clearly demonstrate that, due to SOC, a rotating electric field could produce electron spin-precession that can be viewed as the "reverse spin-electric effect" predicted very recently by Levitov and Rashba[19]. Within the limit of adiabatic approximation the derived formulas are valid regardless of whether eigenvectors of the total Hamiltonian of the problem are explicitly available or not, and are convenient for approximate calculations. The obtained results describe both transverse and longitudinal dynamics of an effective spin, and elucidate the link between the quantum-mechanical, geometrical, and classical pictures of spin rotation. The main formula for the time evolution of the Bloch vector has pure geometric character and is independent of the physical context.

## II. General Framework

Since the pioneering work of Wilczek and Zee[3], it is well recognized that the non-trivial dependence of the degenerate eigenvectors on topological characteristics of the path adiabatically traversed by the system in the parameter space results in non-Abelian holonomic transformations of the initially prepared state. Mead[20] and Segert[21] were the first who demonstrated that adiabatic rotation of an external electric field will lead to non-Abelian gauge effects on degenerate sublevels of a total angular momentum of a paramagnetic atom. The general class of Hamiltonians considered in Refs.[20] and [21] was $H = H_0 + V(t)$, where $H_0$ is a rotationally invariant part of the total Hamiltonian including SOC, $H_{SO}$. Mead and Segert studied systems with strong coupling of spin and orbital degrees of freedom, when anisotropic electric field potential $V$ may be considered as a small perturbation of Russell-Saunders states, $H_{SO} \gg V$, and $J$ is a good quantum number (to zero order in *V*). In Section IV we shall discuss the physical consequences of this approximation. The essentially similar problem has been considered by Zee[22] in context of nuclear quadrupole resonance and by Avron *et al.*[23], who uncovered the fundamental connection between non-Abelian holonomies in T-invariant fermionic systems and quaternionic structure of the Hilbert space.

The theory was extended to cover the case of an arbitrary strong external electric field, which may lead to a complete quenching of the atomic orbital angular momentum, and/or systems with zero electron orbital angular momentum in the ground state[10,11]. In this situation, the SOC is weak ($H_{SO} \ll V$) and it is crucial to transform into the appropriate "adiabatic" basis defined by the eigenvectors of the *static* non-truncated Hamiltonian, $H$, before any calculations of the spin-electric and non-Abelian gauge effects. It has been shown[10,11] that, in the moving (M) frame of reference that follows the rotation of the principal axes of electric field tensor, the evolution of a spinor $\Psi^{(M)}{}_{KD}$ that represents a Kramers doublet, adiabatically isolated from the rest of a spin-multiplet, is governed by the effective spin-Hamiltonian, $H^{(M)}{}_{eff}$:

$$i\dot{\Psi}^{(M)}{}_{KD}(t) = H^{(M)}{}_{eff}(t) \Psi^{(M)}{}_{KD}(t), \qquad (1a)$$

$$H^{(M)}{}_{eff}(t) := -1/2\, \vec{\omega}(t)\, \vec{\gamma}^{(M)}\, \vec{\sigma}^{(M)} = -i A^{(M)}{}_{WZ}(t). \qquad (1b)$$

Here and in the following $\hbar = 1$, $A^{(M)}{}_{WZ}$ is the Wilczek-Zee non-Abelian gauge potential, $\vec{\sigma}^{(M)}$ is the vector of Pauli matrices, and $\vec{\omega}(t)$ is an instantaneous angular velocity of the M-frame relative to the *space-fixed* lab (L) frame at time *t*. The "tensor" $\vec{\gamma}^{(M)}$ is defined by the expression[10,24]

$$1/2\, \vec{\gamma}^{(M)}\, \vec{\sigma}^{(M)} := P^{(M)}{}_{KD}[R^{-1}(t) \vec{J}^{(L)} R(t)] P^{(M)}{}_{KD}, \qquad (2)$$

where $P^{(M)}_{KD}$ is the projector onto the complex two-dimensional (*2D*) Hilbert space spanned by the Kramers doublet (hereafter referred to as projective spinor space). The Schrödinger-type equation (1a) and the expression (1b) depend on a choice of gauge that specifies the *reference* orientation, i.e. the orientation in which the M-frame coincides with some space-fixed frame. At the moment $t = 0$, the reference orientation may always be chosen such that $\vec{\vec{\gamma}}^{(M)}$ is diagonal and that the main axis $Z$ of this "tensor" represents the quantization axis of the effective spin operator $\vec{S}^{(M)}_{eff} := \vec{\sigma}^{(M)}/2$. The rotation operator $R(t)$ in Eq.(2) maps the space-fixed reference orientation into the actual orientation of the M-frame at time $t$, $\Psi^{(M)}(t) = R(t)\Psi^{(L)}(t)$, where $\Psi$ is a true eigenfunction of $H$.

The effective spin-Hamiltonian $H^{(M)}_{eff}$, Eq.(1b), can be viewed as a generic Zeeman Hamiltonian of a spin-1/2 particle in an "effective" time-dependent magnetic field $\vec{\omega}(t)\vec{\vec{\gamma}}^{(M)}$ that appears in the frame that follows the rotation of the principal axes of electric field tensor. The role of the Wilczek-Zee gauge potential becomes clear if we take into account that a differential action of $H^{(M)}_{eff}$ is proportional to the angle of rotation, $|\vec{\omega}(t)|\,dt$, i.e., to the *distance* in the angular space. Correspondingly, $A^{(M)}_{WZ}dt$ provides a pure geometrical mapping between an infinitesimal change in the orientation of the electric field tensor in the real *3D*-Euclidean space and the resultant rotation of $\Psi^{(M)}_{KD}$ in the spinor space. As long as rotation of a Stark field represents an adiabatic perturbation to the system the evolution of the spinor $\Psi^{(M)}_{KD}$ is a unique function of the curve $C$ traversed by the driving electric field in the angular space and is independent of the rate of traversal.

For finite times, the infinitesimal rotations of a spinor accumulate to a finite rotation, thereby giving rise to energy splitting and to transitions between the pair of Kramers-conjugate states. In general, the axis of rotation may change its direction in time, so the elementary rotations of an electric field may not commute. To find the evolution of $\Psi^{(M)}{}_{KD}$ in this situation one must evaluate the path-ordered integral along the segments of $C$ belonging to different coordinate patches and multiply the resulting unitary matrices in the order in which the curve is traversed. The formal solution of Eq.(1a) can be written as a time-ordered product

$$\Psi^{(M)}{}_{KD}(t) = R^{(M)}{}_{eff}(t)\Psi^{(M)}{}_{KD}(0), \quad R^{(M)}{}_{eff}(t) = P\exp[-i\int_0^t H^{(M)}{}_{eff}(t')dt'], \quad (3)$$

where $P$ represents the Dyson operator. The evolution (*rotation*) operator $R^{(M)}{}_{eff}(t)$ in eq.(3) is merely a concise mathematical expression of the composition of the infinitesimal rotations mentioned above[12][13][25].

Note that the original non-truncated Hamiltonian serves only to determine the gauge group and the principal values of $\tilde{\gamma}^{(M)}$. Within the adiabatic limit, i.e., when inverse of $|\vec{\omega}|$ is much larger than the time scale of the quantum system, expression (1b) is valid regardless of whether eigenvectors of $H$ are explicitly available or not, and is convenient for approximate calculations. Clearly, $\tilde{\gamma}^{(M)}$ is not a true tensor, since it does not transform covariantly under a gauge transformation. The explicit form of $P^{(M)}{}_{KD}$ and thus $\tilde{\gamma}^{(M)}$ depends on the problem at hand. For example, in axially symmetric systems $\vec{J}_Z$ commutes with $H$ and (1b) reads[10]:

$$H^{(M)}{}_{eff}(t) = -1/2\{\gamma_\| \vec{\omega}_Z(t)\vec{\sigma}^{(M)}{}_Z + \gamma_\perp[\vec{\omega}_X(t)\vec{\sigma}^{(M)}{}_X + \vec{\omega}_Y(t)\vec{\sigma}^{(M)}{}_Y]\} \quad (4a)$$

$$= -1/2 \{\dot{\alpha}(t) [\gamma_\| \cos\beta(t) \vec{\sigma}^{(M)}{}_Z - \gamma_\perp \sin\beta(t)\, \vec{\sigma}^{(M)}{}_X] + \dot{\beta}(t)\, \gamma_\perp \sigma^{(M)}{}_Y\}. \quad (4b)$$

Here $\gamma_\| := \tilde{\gamma}^{(M)}{}_{ZZ} = 1$, $\gamma_\perp := \tilde{\gamma}^{(M)}{}_{XX} = \tilde{\gamma}^{(M)}{}_{YY}$, $\{\alpha, \beta\}$ is a set of Euler angles specifying the rotation $R(t)$, and the familiar structure of non-Abelian Wilczek-Zee gauge potential[4,22] is easily recognizable in Eq.(4b).

The strength of the surface-induced effective parameter of SOC, $\lambda(1P)$, is of the order of spin-orbit splitting, $E(1P_{3/2}) - E(1P_{1/2})$, between the levels of the first excited electron state $1P$ ($l = 1$) in spherical CdSe nanocrystals. This parameter is size-dependent ($\lambda(1P) \sim 1$ - $10$ meV for QD-radius $40$ - $13$ Å)[18] and is comparable with the fine-structure splitting in alkali atoms. When spin and orbital degrees of freedom are strongly coupled, to zero order in $V$ (see Refs.[20] and [21]), in the $|m_J| = ½$ sector $P^{(M)}{}_{KD} = |J, ½\rangle\langle J, ½| + |J, -½\rangle\langle J, -½|$. Hence,

$$P^{(M)}{}_{KD} \vec{J}^{(M)}{}_Z P^{(M)}{}_{KD} = 1/2\, \vec{\sigma}^{(M)}{}_Z, \quad P^{(M)}{}_{KD} \vec{J}^{(M)}{}_{X;Y} P^{(M)}{}_{KD} = 1/2(J+1/2)\, \vec{\sigma}^{(M)}{}_{X;Y},$$

and correspondingly $\gamma_\| = 1$, $\gamma_\perp = J + 1/2$. These results do not depend on the system being electronic or nuclear. For example, the evolution of the lowest Kramers doublet adiabatically isolated from the rest of a *nuclear* spin-multiplet (half-odd-integer nuclear spin $I > ½$) can be described[22] by Eqs.(1a, b) with $\gamma_\| = 1$, $\gamma_\perp = I + 1/2$.

In the opposite limit, e.g., for systems with zero spatial angular momentum in the ground state and large gap between electron states of different orbital symmetry, SOC is suppressed. As a result, a simple perturbation treatment of the eigenfunctions of the full "unrotated" Hamiltonian $H$ is appropriate. This situation corresponds, e.g., to *s* like electron states in the lowest conduction band $\Gamma_6^c$ in semiconductors and/or paramagnetic complexes of low symmetry, where the orbital angular momentum is quenched by a strong anisotropic "ligand field" potential. Examples of model $\vec{\tilde{\gamma}}$-"tensor" calculations can be found in Refs. [10] and [24]. It has been shown[10] that to first order in SOC

$$\vec{\tilde{\gamma}}^{(M)} - \hat{1} = \Delta\vec{\tilde{g}}^{(M)} \propto \lambda/\Delta E ,$$

where $\Delta\vec{\tilde{g}}^{(M)} = \vec{\tilde{g}}^{(M)} - g_e\hat{1}$ corresponds to the SOC induced deviation of the components of the g-"tensor" from the value of the free electron, $\lambda$ is an effective SOC constant, and $\Delta E$ is the energy separation to the nearest eigenstate of $H$ with different orbital symmetry. Apparently, in the absence of SOC, $\vec{\tilde{\gamma}}^{(M)} = \hat{1}$, spin will be entirely decoupled from the rotation of an external electric field. We shall return to this point in Section IV.

## III.  Coherent rotation

Following the canonical Berry's example (rotation on a cone), consider an electric field that has a constant magnitude and rotates about some fixed axis $\hat{n} = \{\sin\theta\cos\varphi, \sin\theta\sin\varphi, \cos\theta\}$ with a constant frequency $\omega$. In the rotating M-frame the direction of this field relative to the axis of rotation is constant. Thus, within the adiabatic limit, the evolution of a spinor $\Psi^{(M)}{}_{KD}$ is governed by the Schrödinger-type equation (1a) with the *time independent* effective spin-Hamiltonian (1b) and is readily integrable[26][27] (see Eq.(3)):

$$R^{(M)}{}_{eff}(t) = \exp[i(\vec{\omega}t)\hat{\tilde{n}}\vec{\tilde{\gamma}}^{(M)}\vec{S}^{(M)}{}_{eff}] = \cos\phi/2 + i\hat{\tilde{a}}\vec{\sigma}^{(M)}\sin\phi/2. \qquad (5)$$

Here $\phi := (\omega t)A$ denotes the angle of rotation of an effective spin about the axis $\hat{\tilde{a}}$ defined by the unit vector

$$\hat{\tilde{a}} := A^{-1}\hat{\tilde{n}}\vec{\tilde{\gamma}}^{(M)}, \quad A := \|\hat{\tilde{n}}\vec{\tilde{\gamma}}^{(M)}\| = \{\gamma_{ZZ}{}^2\cos^2\theta + [\gamma_{XX}{}^2\cos^2\varphi + \gamma_{YY}{}^2\sin^2\varphi]\sin^2\theta\}^{1/2}. \qquad (6)$$

Utilizing the well-known homomorphism between the vectors and rotation operators of Euclidean space and the spinors and rotation operators of spinor space[26 28 29 30], we may cast the solution of the Schrödinger-type equation (1a) into the following form

$$\vec{u}^{(M)}(t) = \Re^{(M)}_{eff}(\phi, \hat{\vec{a}}) \, \vec{u}^{(M)}(0) = \vec{u}^{(M)} \cos\phi + (\hat{\vec{a}} \times \vec{u}^{(M)})\sin\phi + \hat{\vec{a}}(\hat{\vec{a}} \bullet \vec{u}^{(M)})(1 - \cos\phi). \quad (7)$$

Here we introduce the polarization or Bloch vector $\vec{u}^{(M)}(t) := Tr[\rho^{(M)}_{KD}(t) \vec{\sigma}^{(M)}]$, whose tip traces out the curve $C'$ on the surface of the Bloch sphere, $\Re^{(M)}_{eff}(t)$ stands for the *3D rotation matrix*, $\rho^{(M)}_{KD}(t) := |\Psi^{(M)}_{KD}(t)\rangle\langle\Psi^{(M)}_{KD}(t)|$ is the corresponding density operator, and $\vec{u}^{(M)}$ denotes $\vec{u}^{(M)}(t=0)$. Note the difference between $\hat{\vec{a}}$ and $\hat{\vec{n}}$, $\phi$ and $\omega t$ that reflects the divergence between the curve $C$ and the curve $C'$ on the Bloch sphere.

To perceive the pure geometrical nature of this result decompose vector $\vec{u}^{(M)}(t=0)$ into components parallel $\vec{u}^{(M)}_{\parallel} = \hat{\vec{a}}(\hat{\vec{a}} \bullet \vec{u}^{(M)})$ and perpendicular $\vec{u}^{(M)}_{\perp} = -\hat{\vec{a}} \times (\hat{\vec{a}} \times \vec{u}^{(M)})$ to the rotation axis $\hat{\vec{a}}$. Then, taking into account that only $\vec{u}^{(M)}_{\perp}$ is affected by the rotation about $\hat{\vec{a}}$, obtain

$$\vec{u}^{(M)}(t) = \vec{u}^{(M)}_{\parallel} + \vec{u}^{(M)}_{\perp} = \vec{u}^{(M)}_{\parallel} + \vec{u}^{(M)}_{\perp} \cos\phi + (\hat{\vec{a}} \times \vec{u}^{(M)})\sin\phi. \quad (8)$$

After a minor rearrangement it becomes clear that Eq.(8) is identical to Eq.(7). Thus, we may reiterate that the time evolution of the system under consideration allows a complete geometrical description, i.e. without indicating the schedule of a motion of the electric field tensor in the angular space.

To simplify formulas, without a significant loss of generality, hereafter we shall consider an axially symmetric system and take $\hat{\vec{n}}$ in the XZ plane. Correspondingly $\varphi = 0$, $\hat{\vec{n}} = \{\sin\theta, 0, \cos\theta\}$, $\hat{\vec{a}} = \{\sin\delta, 0, \cos\delta\}$, and

$$\mathfrak{R}^{(M)}{}_{e\!f\!f}(\phi,\hat{\vec{a}}) = \begin{bmatrix} \sin^2\delta + \cos^2\delta\cos\phi & -\sin\phi\cos\delta & 1/2(1-\cos\phi)\sin 2\delta \\ \sin\phi\cos\delta & \cos\phi & -\sin\phi\sin\delta \\ 1/2(1-\cos\phi)\sin 2\delta & \sin\phi\sin\delta & \cos^2\delta + \sin^2\delta\cos\phi \end{bmatrix}, \quad (9)$$

where

$$\sin\delta := (\gamma_\perp/A)\sin\theta, \quad \cos\delta := (\gamma_\parallel/A)\cos\theta, \quad A = [\gamma_\parallel^2\cos^2\theta + \gamma_\perp^2\sin^2\theta]^{1/2}. \quad (10)$$

Note that generally the choice of the *space-fixed* L-frame is defined by the experimental conditions and may not coincide with the reference orientation. For example, it may be convenient to choose the initial state and the detected state of a system to consist of Zeeman magnetization along the rotation axis. In this case, the axis $Z$ of the reference orientation may be tilted at some angle to $Z^{(L)}$ and additional rotational transformations may be required to satisfy the particular choice of experimental conditions. To simplify the notation in the following we shall assume that the reference orientation coincides with the L-frame.

To describe the evolution of a system from an initial state $\vec{u}^{(L)}(0) = \vec{u}^{(M)}(0)$ to a final state $\vec{u}^{(L)}(t)$ in the space-fixed frame we have to transform $\vec{u}^{(M)}(t)$ back to the L-frame. This is merely a reverse rotation of the coordinate system compensating for the rotation of the M-frame. In terms of Euler-Rodrigues parameters[26] the orientation of the M-frame at time $t$ is $\{\cos\omega t/2, \hat{\vec{n}}\sin\omega t/2\}$. Consequently, $\vec{u}^{(L)}(t) = \mathfrak{R}(\omega t, -\hat{\vec{n}})\,\vec{u}^{(M)}(t)$ and the corresponding rotation matrix is

$$\mathfrak{R}(\omega t, -\hat{\vec{n}}) = \begin{bmatrix} \sin^2\theta + \cos^2\theta\cos\omega t & \sin\omega t\cos\theta & 1/2(1-\cos\omega t)\sin 2\theta \\ -\sin\omega t\cos\theta & \cos\omega t & \sin\omega t\sin\theta \\ 1/2(1-\cos\omega t)\sin 2\theta & -\sin\omega t\sin\theta & \cos^2\theta + \sin^2\theta\cos\omega t \end{bmatrix} \quad (11)$$

Thus, the curve traced by the tip of the Bloch vector in the L-frame is derived explicitly

$$\vec{u}^{(L)}(t) = \Re(\omega t, -\hat{n})\Re^{(M)}_{eff}(\phi, \hat{\tilde{a}})\vec{u}^{(M)}(0). \qquad (12)$$

Formula (12) with the aid of Eqs.(2), (9) - (11) is our main analytical result. The evolution of $\vec{u}^{(L)}(t)$ is the resultant of two rotations performed about different axes. The first one is rotation of a Bloch vector through an angle $\phi$ around an effective symmetry axis of the system, $\hat{\tilde{a}} = \{\sin\delta, 0, \cos\delta\}$, that in general case is distinct from $\hat{n}$, and from the symmetry axis $Z$ of the electric field tensor. This rotation is associated with a *physical* change of a state, whereas the second one is rotation around $-\hat{n}$ through an angle ($\omega t$) and is just a coordinate transformation. As mentioned above, Eqs.(9) - (12) have pure geometric character and stand independently of the context: the physical nature of the problem is hidden in the definition of $\tilde{\gamma}^{(M)}$.

## IV. Examples

A). Consider, for example, the rotation of the principal axis $Z$ of electric field tensor about the axis X ($\theta = \pi/2$). It is easy to see from Eq.(10) that in this situation $A = \gamma_\perp$, $\delta = \theta = \pi/2$, and $\phi = (\omega t)\gamma_\perp$. As a result, the Bloch vector and the Stark field are revolving about the same axis, and Eq.(12) reads

$$\vec{u}^{(L)}(t) = \begin{bmatrix} 1 & 0 & 0 \\ 0 & \cos[\omega t(\gamma_\perp - 1)] & -\sin[\omega t(\gamma_\perp - 1)] \\ 0 & \sin[\omega t(\gamma_\perp - 1)] & \cos[\omega t(\gamma_\perp - 1)] \end{bmatrix} \vec{u}^{(M)}(0) \qquad (13)$$

Let the polarization vector be initially directed along the principal axis of the external electric field tensor, $\vec{u}^{(L)}(0) = \vec{u}^{(M)}(0) = \{0, 0, 1\}$. In this situation, formula (13) predicts the reversal of a sign of the component $\vec{u}^{(L)}{}_z(t)$, along its original direction in space $\vec{u}^{(L)}{}_z(0)$, when the driving electric field swept the angle $\omega t_R = \pi/(\gamma_\perp - 1)$. Furthermore, $\vec{u}^{(L)}(t)$ returns to its initial orientation after $k := 1/(\gamma_\perp - 1)$ complete turns of a field. The evolution loop here is not related to the loop in the parameter space, but rather to the projective spinor space as in the Aharonov-Anandan analysis[31]. In fact, Eq.(13) clearly demonstrates that a rotating electric field could produce oscillations of an effective spin-½, similar to Rabi precession in the external magnetic field.

For particles with strong SOC ($H_{SO} \gg V$) "non-centralness"[32] of the system is small and to zero order in the external electric field potential Eq.(13) can be reduced to the result obtained in the M-frame by Mead ($\gamma_\perp = J + ½$)[20]. Note that within this approximation, for systems with $J = ½$, e.g., the first excited electron state $1^2P_{1/2}$ in spherical CdSe nanocrystals[18], $\vec{J}$ will be *completely* decoupled from the adiabatic rotation of an external electric field, $\gamma_\perp = 1$, $k \to \infty$, and $\vec{u}^{(L)}(t) = \vec{u}^{(L)}(0)$ at all times. This is the result of approximate calculations made in Refs.[20] and [21]. To zero order in $V$, at each fixed orientation of a Stark field $|J\ m_J\rangle$ comprise the adiabatic eigenbasis of the instantaneous Hamiltonian. Hence, in accordance with Kramers' theorem, the amplitude of spin-flip transitions between the components $|m_J| = ½$ ($J = ½$) induced by an electric field is negligible. The same approximation, however, may lead to the non-trivial results[20,21,22] if the system under study is in the $J > ½$ multiplet. An external Stark field may split levels and induce transitions between the states differing in $J$-projection (e.g., $|m_J| = 1/2$ and $|m_J| = 3/2$). It is easy to see from Eq.(13) that for $J = 3/2$ (in the $|m_J| = ½$ sector) the Bloch vector $\vec{u}^{(L)}{}_Z(t)$ just follows the Z principal axis of the electric field tensor. For $J > 3/2$ the polarization vector rotates a bit slower than the field.

In the opposite limit, for weak coupling of spin and orbital degrees of freedom ($H_{SO} << V$), stationary wave functions of $H_0$ are deformed by a Stark field, "non-centralness" of the system is large, and, as mentioned in Section II, it is necessary to include $V$ in the static non-truncated Hamiltonian[10]. Obviously, if we totally ignore SOC (more accurately all relativistic effects[33]), $\gamma_\perp = 1$, spin will be completely decoupled from the rotation of an external electric field ($k \to \infty$). However, even a small spin-orbit interaction will violate the spin rotation symmetry and one must consider an effective angular momentum of a mixed spin and orbital nature. Since $V$ is included in $H$, the corresponding eigenbasis will follow the adiabatic rotation of the axes of electric field tensor. Consequently, $\vec{u}^{(L)}{}_Z(t)$ will go after the revolving filed, yet, due to the weakness of SOC, it will be left far behind the race of the field ($\Delta g_\perp = |\gamma_\perp - 1| << 1, k >> 1$).

B). Now let $\theta$ be arbitrary. After a time interval $T = 2\pi/\omega$ the electric field tensor returns to the reference orientation, $\Re(T) = 1$, generating the rotation $\phi = 2\pi A$ of a Bloch vector, $\vec{u}^{(L)}(T) = \vec{u}^{(M)}(T) = \Re^{(M)}{}_{eff}(T) \vec{u}^{(M)}(0)$. Assume, once again, that the polarization vector is initially parallel to $Z$. It is easy to see from Eq.(9) and Eq.(12) that in this case

$$\vec{u}^{(L)}{}_Z(T) = \vec{u}^{(M)}{}_Z(T) = \cos^2\delta + \sin^2\delta \cos\phi = 1 - 2\sin^2\delta \sin^2(\pi A). \qquad (14)$$

Correspondingly, the probability of spin-flip transitions is $P(T) = \sin^2\delta \sin^2(\pi A)$. For systems with strong SOC, Eq.(14) can be reduced to results obtained in Refs.[20]-[22]. As expected, this formula has clear geometrical roots. Indeed, let $\alpha$ be the angle between $\vec{u}^{(M)}(t)$ and $\vec{u}^{(M)}(0)$, then $\vec{u}^{(M)}{}_Z(t) = \cos\alpha$, and, with a little help of spherical trigonometry, Eq.(14) can be derived from the simple geometry of the problem[34].

It is interesting to observe in this example the existence of cyclic solutions in the rotating frame of reference. Indeed, from Eq.(9) and Eq.(12), for $\vec{u}^{(M)}(0) = \{0, 0, 1\}$, we have

$$\vec{u}^{(M)}{}_X(t) = \sin 2\delta \sin^2(\omega t A/2),$$

$$\vec{u}^{(M)}{}_Y(t) = -\sin\delta \sin(\omega t A),$$

$$\vec{u}^{(M)}{}_Z(t) = 1 - 2\sin^2\delta \sin^2(\omega t A/2). \qquad (15)$$

Hence, in the M-frame one may achieve a loop on the Bloch sphere when the Stark field completes the angle $\omega t_C = 2\pi/A$.

C). In order to explore the non-Abelian character of the quantum evolution we have to alter the simple orbital motion of the electric field considered above, e.g., allow an adiabatic change of the axis of rotation $\hat{n}$ in time. In this situation it would be necessary to compute the path-ordered integral in Eq.(3) along the *noncommuting* segments of $C$. In general, this task can be accomplished only numerically. To obtain the analytical solution in the following we permit only two orientations of the rotation axis. To elucidate the non-Abelian holonomy we will consider the figure-8 path, i.e., two cones of identical apex angles $\theta$ sharing the line between the vertex and the central point of the figure, $\hat{n}_1 = \{\sin\theta, 0, \cos\theta\}$, $\hat{n}_2 = \{-\sin\theta, 0, \cos\theta\}$. If a driving electric field traverses these two cones in opposite directions, the net solid angle this trajectory will subtend at the center of the figure-8 is zero. As a result, any Abelian effects would naturally cancel out[35][36].

To calculate the motion of the Bloch vector for two adiabatically completed loops, described above, one must evaluate and multiply the resulting unitary matrices in the order in which the curve is traversed. Assume that the reference orientation of the M-frame coincides with the L-frame with $Z^{(L)}$ along the line connecting the vertex and central point of the figure-8. Let the principal axis $Z$ of the electric field tensor traverse both cones counter-clockwise with the constant frequency ω, which corresponds to the opposite sense of rotations relative to $Z^{(L)}$. The resultant motion of the Bloch vector is given by

$$\vec{u}^{(M)}(t) = \Re^{(M)}{}_{2\,eff}(\phi_2, \hat{\vec{a}}_2)\, \Re^{(M)}{}_{1\,eff}(\phi_1, \hat{\vec{a}}_1)\vec{u}^{(M)}(0) \quad, \tag{16}$$

where $\phi_{1,2} = 2\pi A$, $\hat{\vec{a}}_{1,2} = \hat{\vec{n}}_{1,2}\vec{\gamma}^{(M)}/A$. It is straightforward to show that after the completion of both loops, for $\vec{u}^{(L)}(0) = \vec{u}^{(M)}(0) = \{0, 0, 1\}$, we have

$$\vec{u}^{(L)}{}_Z(2T) = \vec{u}^{(M)}{}_Z(2T) = 1 - 2\sin^2 2\delta \, \sin^2 \pi A \, (1 + 2\cos^2 \pi A). \tag{17}$$

Thus, the non-Abelian holonomy leads to a reversal of the direction of the Bloch vector with the probability $P(2T) = \sin^2 2\delta \, \sin^2 \pi A \, (1 + 2\cos^2 \pi A)$. As expected, *P(2T)* disappears *only* if the rotation of the electric field about the axis -X ($\theta = -\pi/2$) immediately follows the rotation about the axis X ($\theta = \pi/2$). In this case, $[\Re^{(M)}{}_{2\,eff}(t), \Re^{(M)}{}_{1\,eff}(t)] = 0$, and *Abelian* holonomies that accumulated during the two halves of the figure-8 loop cancel each other out.

Finally, let the principal axis $Z$ of the electric field tensor traverse the first cone counter-clockwise and the second one clockwise ($\phi_1 = 2\pi A = -\phi_2$). This situation corresponds to the same sense of rotations relative to $Z^{(L)}$. As a result

$$\vec{u}^{(L)}{}_Z(2T) = \vec{u}^{(M)}{}_Z(2T) = 1 - 2(\sin^2 2\delta \, \sin^2 \pi A + \sin^4 \delta \, \sin^2 2\pi A), \tag{18}$$

and $P(2T) = 4\sin^2\delta \sin^2\pi A (\cos^2\delta + \sin^2\delta \cos^2\pi A)$. Apparently, in the Abelian case ($\theta = \pi/2$) Eq.(18) reduces to Eq.(14) with $t = 2T$, as it should, since this situation corresponds to two consecutive loops made in the same direction.

## V.  Summary

The response of an *effective* spin to the adiabatic perturbation induced by the revolving electric field of constant frequency and magnitude has been calculated exactly. The matrices in the main formula, Eq.(12), for the time evolution of the Bloch vector correspond to simple geometrical operations that help to uncover the clear geometrical picture of the obtained results. The relation expressed in Eq.(7) simply describes a *3D-rotation* of the Bloch vector in the frame rotating synchronously with the driving electric field. It merely represents the covariant rotational transformation of a *classical* vector, which, regardless of the context, is essentially a pure geometric action. In fact, in *rigid-body* mechanics formula (7) is well known as Rodrigues' formula[26]. In the context of the canonical Berry's problem, it has been re-derived very recently in[37] and is implicitly present in its matrix form in[38].

Within the limit of adiabatic approximation the derived formulas are valid regardless of whether eigenvectors of the total Hamiltonian of the problem are explicitly available or not, and are convenient for approximate calculations. The main expression for the time evolution of the Bloch vector has pure geometric character and is independent of the physical context.

I whish to thank Prof. Steiner for numerous stimulating discussions as well as financial support of Alexander von Humboldt Foundation which made possible for me to work in Prof. Steiner's research group at the University of Konstanz.

[1] M. V. Berry, Proc. R. Soc. London Ser. A **392**, 45 (1984).

[2] *Geometric Phase in Physics*, edited by A. Shapere and F. Wilczek (World Scientific, Singapore, 1989).

[3] F. Wilczek and A Zee, Phys. Rev. Lett. **52**, 2111 (1984).

[4] P. Solinas *et al.,* Phys. Rev. A 67, 062315 (2003).

[5] G. Salis *et al.*, Nature (London) **414**, 619 (2001).

[6] Y. Kato *et al.*, Science **299**, 1201 (2003).

[7] E. I. Rashba and Al. L. Efros, Phys. Rev. Lett. **91**, 126405 (2003).

[8] J. H. Van Vleck, Phys. Rev. **57**, 426 (1940).

[9] D. Pines, J. Bardeen, and C. Slichter, Phys. Rev. **106**, 489 (1957).

[10] U. E. Steiner and Yu. A. Serebrennikov, J. Chem. Phys. **100**, 7503 (1994).

[11] Yu. A. Serebrennikov and U. E. Steiner, Mol. Phys. **84**, 627 (1995).

[12] Yu. A. Serebrennikov and U. E. Steiner, J. Chem. Phys **100**, 7508 (1994).

[13] Yu. A. Serebrennikov and U. E. Steiner, Chem. Phys. Lett. **222**, 309 (1994).

[14] A. V. Khaetskii and Yu. V. Nazarov, Phys. Rev. B **64**, 125316 (2001).

[15] L. M. Woods *et al.*, Phys. Rev. B **66**, 161318 (2002).

[16] C. Tahan *et al.*, Phys. Rev. B **66**, 035314 (2002).

[17] B. A. Glavin and K.W. Kim, Phys. Rev. B **68**, 045308 (2003).

[18] A. V. Rodina, Al. L. Efros, and A. Yu. Alekseev, Phys. Rev. B **67**, 155312 (2003).

[19] L. S. Levitov and E. I. Rashba, Phys. Rev. B **67**, 115324 (2003).

[20] C. A. Mead, Phys. Rev. Lett. **59**, 161 (1987).

[21] J. Segert, J. Math. Phys. **28**, 2102 (1987).

[22] A. Zee, Phys. Rev. A **38**, 1 (1988).

[23] J. E. Avron, L. Sadun, J. Segert, and B. Simon, Phys. Rev. Lett. **61**, 1329 (1988).

[24] U. E. Steiner and D. Bursner, Z. Phys. Chem. N.F. **169**, 159 (1990).

[25] In Ref.12, the theory of electron spin-rotational relaxation was formulated in terms of stochastic modulation of non-Abelian gauge potential. Note that the key approximation made in Ref.[12] - equation (22) - that leads to equation (21) obviously should precede this equation. In the text of the article it erroneously follows eq.(21).

[26] S. L. Altmann, *Rotations, Quaternions and Double Groups* (Clarendon Press, Oxford, 1986)

[27] D. A. Varshalovich *et al.*, *Quantum Theory of Angular Momentum* (World Scientific, Singapore, 1988).

[28] G. Town and D. Rosenfeld, Phys. Rev. A **40**, 3429 (1989).

[29] D. Fernandez, L. M. Nieto, M. A. Olmo, and M. Santader, J. Phys. A **25**, 5151 (1992).

[30] S-L. Zhu, Z. D. Wang, and Y-D.Zhang, Phys. Rev. B **61**, 1142 (2000).

[31] Y. Aharonov and J. Anandan, Phys. Rev.Lett. **58**, 1593 (1987).

[32] J. H. Van Vleck, Phys. Rev. **33**, 467 (1929).

[33] B. Mashhoon, Phys. Rev. Lett. **61**, 2639 (1988).

[34] I. I. Rabi, N. F. Ramsey, and J. Schwinger, Rev. Mod. Phys. **26**, 167 (1954).

[35] S. Appelt, G. Wackerle, and M. Mehring, Phys. Lett. A **204**, 210 (1995).

[36] D. P. Arovas and Y. Lyanda-Geller, Phys. Rev. B **57**, 12302 (1998).

[37] Q-G. Lin, J. Phys. A **35**, 377 (2002).

[38] S-L. Zhu and Z. D. Wang, Phys. Rev. Lett. **85**, 1076 (2000).